\documentclass[12pt]{article}

\def\PI{$\pi $}
\def\Af{$\alpha\ $}
\def\HH{$\hbar $}
\def\E{$e$}
\def\EE{$e^2$}
\def\C{$c$}
\def\AF#1{\alpha_#1}
\def\AFr{\AF\rho}
\def\eq{&\!\!\!=\!\!&}
\def\deq{&\!\!:=\!\!&}

\begin{document}

\title{DETERMINATION OF THE FINE STRUCTURE CONSTANT BY {\huge $\pi $}}
\author{A. H. Anastassov \\
\date{October 31, 1997}
University of Sofia, Faculty of Physics \footnote{petersl@phys.uni-sofia.bg}}
\maketitle
\begin{abstract}
In the present report the author presents a simple and systematically
defined formula for the fine structure constant based only on the
number \PI.  The
difference between the suggested value and the currently known experimental
one is about 60 times smaller than the uncertainty interval.
\end{abstract}
\bigskip \ The fundamental physical constants have attracted the attention of
physicists and mathematicians for long time. Especially challenging is the
idea that they can be reduced to mathematical constants (Hilbert \cite{Hb},
Einstein \cite{Es}).

The greatest is the challenge offered by the dimensionless constant
\Af of Sommerfeld \cite{Sm}, currently known as the fine structure constant
or as electromagnetic constant. As it is well known, it is determined by the
dimensionless relation \EE/\HH\C\  , where \E{\,}\ is the
elementary electric charge, \HH\ is Plank's constant divided by 2\PI ,
and \C\ - the absolute velocity (the speed of light).
Its reciprocal value 1/\Af is very closed to the integer $137$ and this fact
has grown the attention of many researches starting with Sommerfeld himself.
Eddington's special interest towards this issue is well known, too \cite{Ed}.
Similarly Dirac assumed that some constants, including the fine structure
constant \Af , where build up of simple quantities like
4\PI\ \cite{Di}. In this line of research were published
interesting formulae such as:

\begin{eqnarray*}
(2^45!/\pi ^5)^{1/4}(8\pi ^4/9) \eq 137,036\,082 \\
108\pi (8/1843)^{1/6} \eq 137,035\,915 \\
4\pi ^5/9+37/36 \eq 137,036\,527 \\
(137^2+\pi ^2)^{1/2} \eq 137,036\,016
\end{eqnarray*}
considered in more details in \cite{Eg}.

The currently recommended by The National Institute of Standards and
Technology (USA) value of Sommerfeld's constant and its reciprocal are
\cite{Co}:
\begin{eqnarray}\;\;\;\;\;\;\;\;\;\;\;
\label{e1}
\alpha_R \eq {\bf 7,297\,353}\,08(33) \times 10^{-3} \;, \\
\label{e2}
\alpha_R^{-1}\! \eq {\bf 137,035\,98}9\,5(61)\; .
\end{eqnarray}
In this note, following Dirac's idea I will show that \Af could be presented
with an absolute (for the moment) accuracy in \textbf{a simple and
systematical way only through the numbers 1, 2, 3, 4 and \PI }.
In order to do that I will introduce the quantities:
\begin{eqnarray*}
\AF0\!\deq 1/\sqrt{\frac 12\left(\,\frac 12\!+\!1\!\right)}(4\pi)^2 , \\
\rho\deq \AF0/2\pi\!\left( \pi ^2\!-\!2\pi \right) , \\
  L \deq 2\pi \rho , \\
  S \deq \pi \rho ^2, \\
  V \deq \frac 43\pi\rho^3.
\end{eqnarray*}
Proposition: the number
\begin{eqnarray}\ \ \ \ \ \ \ \ \ \ \,
\nonumber
\AFr\,:=\:\AF0/\left( 1+L-2S+3\,V/2\right)
\end{eqnarray}
has the value
\begin{equation}
\label{e4}
\;\;\;\;\;\;\;\AFr\,=\: {\bf 7,297\,353\,085\,4}...\times 10^{-3} \\
\end{equation}
and its reciprocal value is
\begin{equation}
\label{e5}
\!\!\!{\AFr^{-1}\!=\,{\bf 137,035\,989\,392}...}\; .
\end{equation}

The comparison shows that the number (\ref{e4}) not only is in the limits of the
error of the experimental value (\ref{e1}) but hits its central value. The same is
valid, of course, for the reciprocal values (\ref{e5}) and (\ref{e2}).

It comes to prove that Sommerfeld's constant can in fact be presented
through simple mathematical quantities, based on the number \PI .


\end{document}